\documentclass[aps,pre,twocolumn,groupedaddress]{revtex4-1}

\usepackage{graphicx}

\newcommand{\eqn}[1]{
\begin{eqnarray}
	#1
\end{eqnarray}
}

\begin{document}


\title{General achievable bound of extractable work under feedback control}


\author{Yuto Ashida}
\author{Ken Funo}
\author{Y\^uto Murashita}
\author{Masahito Ueda}
\affiliation{Department of Physics, University of Tokyo, 7-3-1 Hongo, Bunkyo-ku, Tokyo
113-8654, Japan}


\date{\today}

\begin{abstract}
A general achievable upper bound of extractable work under feedback control is given, where nonequilibrium equalities are generalized so as to be applicable to error-free measurements. The upper bound involves a term which arises from the part of the process whose information becomes  unavailable and is related to the weight of the singular part of the reference probability measure. The obtained upper bound of extractable work is more stringent than the hitherto known one and sets a general achievable bound for a given feedback protocol. Guiding principles of designing the optimal protocol are also suggested.  Examples are presented to illustrate our general results.
\end{abstract}


\maketitle

\section{Introduction}
 Our understanding of nonequilibrium statistical mechanics has significantly deepened over the last couple of decades, in large part owing to the discovery of nonequilibrium relations \cite{ECM93,GC95,Kur98,LS99,Mae99,Jar97Apr,Jar00,Cro98,Cro99,HS01,KpvB07} and rapid advances in experimental techniques \cite{WSMe02,LDSe02,DCPR05,CRJe05,LBK08,TSUe10}. Against such a backdrop, there has  been a renewed interest in the relationship between information and thermodynamics \cite{KQ07,ADG08,EvB11,SU08,SU09,SU10,HVS10,HP11,Kim11,AS11,AS12,ES12,SU12}. In particular, feedback control \cite{LBK08,TSUe10,SU08,SU09,SU10,HVS10,HP11,Kim11,AS11,AS12,ES12,SU12} provides a key framework for understanding the role of Maxwell's demon \cite{Max1871,LM10,MNV09}. In modern terms, Maxwell's demon is a feedback controller who performs measurement on a system and utilizes the obtained information \cite{MNV09} to extract work from the system beyond the limit set by the conventional second law of thermodynamics  \cite{SU08,SU09,SU10}: 
 \eqn{\label{ineq1}
  -\langle W_{\rm{d}} \rangle \leq k_{\rm{B}}T\langle I \rangle,
 }
 where $W_{\rm{d}}=W-\Delta F$ is the dissipated work which is defined as the performed work $W$ minus the associated free energy difference $\Delta F$,  and $I$ is the amount of information that the demon acquires by measurements. The inequality (\ref{ineq1}) implies that feedback control enables one to extract work up to the acquired information multiplied by $k_{\rm{B}}T$, where $k_{\rm B}$ is the Boltzmann constant and $T$ is the absolute temperature.
 
   One can, in principle, construct a protocol that achieves the equality in (\ref{ineq1}) \cite{Ja09,HITD10,THD10,EvB11}. However, an unlimited number of external parameters are required to do so, which presents a daunting challenge for experimenters. In practice, we often encounter a situation in which the right-hand side of the inequality (\ref{ineq1}) does not give an achievable bound of extractable work \cite{LBK08,TSUe10}. Hence it is natural to ask the following question: what is the fundamental achievable upper bound of extractable work by means of a given feedback control?  Since  a feedback protocol is normally designed to work best  under error-free measurements, extractable work will be maximized when measurements are performed perfectly. In this respect, we need to consider systems under error-free measurements to answer the above question and therefore, it is essential to work with the framework which allows us to derive nonequilibrium equalities under error-free measurements. However, it is known that there is a difficulty to formulate nonequilibrium equalities under error-free measurements as discussed in Refs. \cite{SU12,HVS10,MFU14}. This is because error-free measurements project the post-measurement state into a sharply localized region in phase space and the forward path probability vanishes outside of this region; nevertheless, the backward path can penetrate into such  projected-out regions due to thermal fluctuations. The known nonequilibrium equalities implicitly assume the absence of those paths.
   
   \begin{figure}[b]
\includegraphics[scale=0.4]{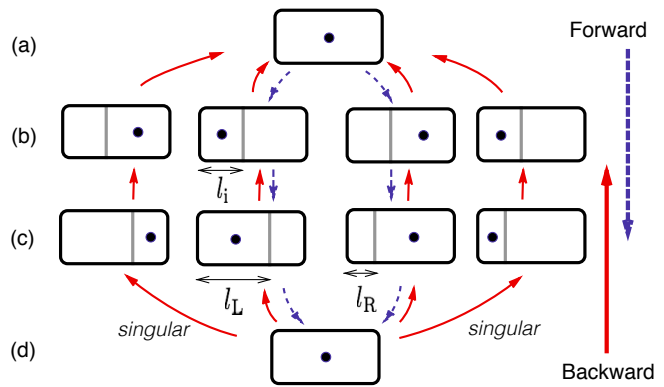}
\caption{\label{figS} (Color online) A single-particle classical Szilard engine. The forward (backward) protocol is indicated by dashed downward (solid upward) arrows. (a) The system is initially in thermal equilibrium at temperature $T$. (b) A barrier is inserted at position $l_{\rm i}$ and an error-free measurement is performed to find in which partition (left (L) or right (R)) the particle resides. (c) Depending on the measurement outcome L or R, the barrier is moved to the position $l_{\rm L}$ or $l_{\rm R}$. (d) The barrier is removed and the system becomes thermalized at temperature $T$. A certain amount of information becomes unavailable upon the removal of the barrier from (c) to (d). The backward process (solid upward arrows) starts at the equilibrium state in (d). The barrier is inserted at $l_{\rm L,R}$ depending on the measurement outcome of the forward process in (c). Then the barrier is moved to $l_{\rm i}$ in (b) and finally it is removed in (a). The leftmost (rightmost) path is a singular one which does not exist in the forward path corresponding to the measurement outcome L (R).}
\end{figure}

   To elucidate this point, let us consider a simple generalization of the Szilard engine (as illustrated in Fig. \ref{figS}).  In this model, the measurements are assumed to be error-free so that a particle resides in the left (right) partition if and only if  the measurement outcome is left (L) (right (R)). Thus, the forward path, which is indicated by dashed downward arrows,  involves only two cases in which the particle resides in the left (right) partition if the measurement outcome is L (R). Nevertheless, the backward path (indicated by solid upward arrows) involves the additional two cases in which the particle resides in the right (left) partition for the measurement outcome L (R) due to thermal diffusion of the particle. These two cases are shown as the leftmost and rightmost paths in Fig. \ref{figS} and are referred as  {\it singular} paths. The known nonequilibrium equalities are based on the assumption that  no such singular paths exist. Therefore, the known derivations of nonequilibrium equalities \cite{SU10,AS12,SJ12} are not applicable to the above situation. This is the difficulty one encounters when one considers systems under error-free measurements.
  
  In this paper, we formulate nonequilibrium equalities under error-free measurements and feedback control by identifying the origin of the above difficulty as some pieces of information  becoming  unavailable in the feedback process. The main result is Eq. (\ref{FT2}) in which a new term $I_{\rm{u}}$ quantifies the amount of information that becomes unavailable in the feedback process. Such a loss of information occurs due to thermodynamic irreversibilities inherent in the feedback process, {\it e.g.}, relaxation and free expansions.  The equality (\ref{FT2}) leads to the inequality (\ref{sec}) which gives a general achievable bound of extractable work for a given feedback control. The new term $I_{\rm u}$ reduces  the achievable bound of extractable work and leads to a more stringent inequality than the known one (\ref{ineq1}). We discuss a few examples in which our inequality (\ref{sec}) gives an achievable upper bound of extractable work whereas the inequality (\ref{ineq1}) does not.

This paper is organized as follows. In Sec. \ref{noneq}, we discuss the effect of unavailable information on nonequilibrium processes  and derive the main result (\ref{FT2}). The physical implications of the main result are also discussed. In Sec. \ref{ex}, we illustrate our general results by discussing three examples: a Brownian particle in a harmonic potential, and classical and quantum Szilard engines. In Sec. \ref{con}, we conclude this paper.

\section{Nonequilibrium Equality \label{noneq}}
\subsection{Feedback Control}
We consider a stochastic thermodynamic system subject to  error-free measurements and  repeated discrete feedback control. The protocol is performed from $t=0$ to $t=\tau$ and the measurements are performed $N$ times at $t_{i}$ ($i=1,2,\cdots,N$), where  $0<t_{1}<t_{2}<\cdots<t_{N}<\tau$. Let $m_{k}$ be the outcome of the measurement performed at $t=t_{k}$. We denote a region in phase-space corresponding to the outcome $m_{k}$ by $X_{m_{k}}$; the outcome is $m_{k}$ if and only if the phase space point of the system satisfies $\Gamma_{t_{k}}\in X_{m_{k}}$. Because the measurements are error-free, we assume that for each $k$  the set of $X_{m_{k}}$ constitutes a non-overlapping set covering the entire phase space. Since the system is subject to a feedback control, the control parameters $\lambda(t)$ depend on the measurement outcomes up to time $t$ and we denote $\lambda\bigl(t;\{m^{k}\}\bigr)$ if $t_{k}<t<t_{k+1}$. Here $\{m^{k}\}$ is the path of the measurement outcomes up to $k$, {\it i.e.}, $\{m^{k}\}\equiv(m_{1},m_{2},...,m_{k})$. The full protocol executed from $t=0$ to $t=\tau$ is represented by $\Lambda \bigl(\{m^{N}\} \bigr)\equiv\Bigl \{\lambda(t),\lambda\bigl(t;\{m^{1}\}\bigr),...,\lambda\bigl(t;\{m^{N}\}\bigr)\Bigr \}$.

Let $\mathcal{I}_{k}$ be the information acquired by the measurement performed at $t=t_{k}$, which is given by a change in the uncertainty of our knowledge about the system upon making the measurement:
\eqn{
\mathcal{I}_{k}=-\ln p\bigl (m_{k}|\{m^{k-1}\}\bigr ),
}
where $p\bigl (m_{k}|\{m^{k-1}\}\bigr )$ is the conditional probability of obtaining $m_{k}$ on condition that the previous outcome is $\{m^{k-1}\}$. The total amount of the acquired information associated with the measurement outcomes $\{m^{N}\}$ is 
\eqn{\label{Sh}
I =\sum_{k=1}^{N} \mathcal{I}_{k}&=&-\ln\biggl[\prod_{k=1}^{N}p\bigl (m_{k}|\{m^{k-1}\}\bigr )\biggr]\nonumber \\
&=&-\ln P\bigl ( \{m^{N}\}\bigr ),
}
where $P\bigl ( \{m^{N}\}\bigr )$ is the probability of obtaining the measurement outcomes $\{m^{N}\}$ and the third equality follows from Bayes' theorem. Note that since the measurement is error-free, Eq. (\ref{Sh}) is equal to the Shannon entropy rather than the mutual information \cite{SU10,AS12,SJ12}.
\subsection{Unavailable Information}
For each realization of the forward process $\Lambda \bigl(\{m^{N}\} \bigr)$, we consider the associated backward process as follows. We first prepare the system to be in an equilibrium state for a given set of external parameters $\lambda\bigl(\tau;\{m^{N}\}\bigr)$, which is the same as the set of the final values of the external parameters in the forward process. Then the external parameters are varied according to the time-reversed protocol as $\lambda^{\rm{R}}_{k}(t)=\lambda\bigl(\tau-t;\{m^{k}\}\bigr)$ if $\tau-t_{k+1}<t<\tau-t_{k}$. The full backward protocol is parametrized as $\Lambda^{\rm{R}} \bigl(\{m^{N}\} \bigr)\equiv\Bigl \{\lambda^{\rm{R}}_{N}(t),\lambda^{\rm{R}}_{N-1}(t),...,\lambda^{\rm{R}}_{0}(t)\Bigr \}$. Note that no feedback control is performed in the backward process because here the backward process is introduced to quantify the irreversibility of the forward feedback process \cite{HVS10}. 

Let us assume that we perform measurements in the backward process without feedback control. The probability of outcome $m^{\dagger}_{k}$ being observed at $t=\tau-t_{k}$ depends on the history of the outcomes $\overline{\{m^{\dagger k}\}}$, where $\overline{\{m^{k}\}}\equiv(m_{k+1},m_{k+2},...,m_{N})$ and $m^{\dagger}_{k}$ indicates the outcome corresponding to the phase-space region $X^{\dagger}_{m_{k}}$ which is given by  $X_{m_{k}}$ with the sign of the momentum inverted. Let $\rho^{\rm{R}}\Bigl(m^{\dagger}_{k}{\Big|}\overline{\{m^{\dagger k}\}},\Lambda^{\rm{R}} \bigl(\{m^{N}\} \bigr)\Bigr)$ be the conditional probability of outcome $m^{\dagger}_{k}$ being observed on condition that we performed the backward protocol $\Lambda^{\rm{R}} \bigl(\{m^{N}\} \bigr)$ and obtained outcomes $\overline{\{m^{\dagger k}\}}$ at earlier times. We introduce a quantity, which we call unavailable information, as follows:
\eqn{\label{unavb}
\mathcal{I}_{{\rm u},k}\equiv-\ln\rho^{\rm R}\Bigl(m^{\dagger}_{k}\Big|\overline{\{m^{\dagger k}\}},\Lambda^{\rm R} \bigl(\{m^{N}\} \bigr)\Bigr).
}
When all the backward paths return to the region observed by the measurements, $\mathcal{I}_{{\rm u},k}$ vanishes. On the other hand, when there is a fraction of the paths going outside of the observed region, $\mathcal{I}_{{\rm u},k}$ takes on a non-zero value and hence,  the unavailable information quantifies the intrinsic irreversibility of the feedback protocol. Furthermore, the total amount of unavailable information is 
\eqn{
I_{\rm{u}}&=&\sum_{k=1}^{N}\mathcal{I}_{{\rm u},k}\nonumber\\
&=&-\ln\biggl[\prod_{k=1}^{N}\rho^{\rm R}\Bigl(m^{\dagger}_{k}\Big|\overline{\{m^{\dagger k}\}},\Lambda^{\rm R} \bigl(\{m^{N}\} \bigr)\Bigr)\biggr]\nonumber\\
&=&-\ln\rho^{\rm R}\Bigl(\{m^{\dagger N}\}\Big|\Lambda^{\rm R} \bigl(\{m^{N}\} \bigr)\Bigr)\label{unav},
}
where $\rho^{\rm R}\Bigl(\{m^{\dagger N}\}\Big|\Lambda^{\rm R} \bigl(\{m^{N}\} \bigr)\Bigr)$ is the conditional probability of obtaining the set of measurement outcomes $\{m^{\dagger N}\}$ on condition that the backward protocol is $\Lambda^{\rm R} \bigl(\{m^{N}\} \bigr)$. Physically, the unavailable information represents the amount of information that becomes unavailable for use in extracting work in the  feedback process as discussed in detail in Sec. \ref{ex}. 

We can make another interpretation of the unavailable information in terms of the weight of the singular part of the reference measure \cite{MFU14}. Since error-free measurements project out the post-measurement state into a sharply localized region, there should exist those paths which are observed in the backward process but excluded in the forward process. These paths result in measurement outcomes in the backward process which are different from those in the forward process. Such contributions are mathematically  classified as a singular part according to Lebesgue's decomposition theorem in measure theory. Let us consider the weight of the singular part $\lambda_{{\rm S},\{m^{N}\}}$ associated with a realization $\{m^{N}\}$. This is equal to the probability of observing such outcomes in the backward process that are absent from the forward process:
\eqn{\label{sing}
\lambda_{{\rm S},\{m^{N}\}}=1-\rho^{\rm R}\Bigl(\{m^{\dagger N}\}\Big|\Lambda^{\rm R} \bigl(\{m^{N}\} \bigr)\Bigr).
}
From Eqs. (\ref{unav}) and (\ref{sing}) we obtain an alternative expression of the unavailable information 
\eqn{\label{unavsin}
I_{\rm{u}}=-\ln\bigl(1-\lambda_{{\rm S},\{m^{N}\}}\bigr).
}
Thus, the singular part of the probability measure determines the unavailable information. The unavailable information results in a decrease in the achievable upper bound of extractable work as shown below.
\subsection{Nonequilibrium Equality}
We are now in a position to derive nonequilibrium equalities which are applicable to systems subject to error-free measurements. Our goal is to derive the equality (\ref{FT2}) and the inequality (\ref{sec}), the latter giving a general achievable upper bound of extractable work.

We derive nonequilibrium equalities on the basis of the framework of stochastic thermodynamics \cite{HS07,Sei12}. To do so, we consider the reference probability density $\mathcal{P}^{\rm r}[\Gamma^{\rm r}]$ and define the entropy production as
\eqn{
\sigma[\Gamma_{t}]=\ln\frac{\mathcal{P}[\Gamma_{t}]}{\mathcal{P}^{\rm r}[\Gamma^{\rm r}_{t}]}=\ln \frac{p(\Gamma_{0})}{p^{\rm r}(\Gamma^{\rm r}_{0})}+\ln\frac{\mathcal{P}[\Gamma_{t}|\Gamma_{0}]}{\mathcal{P}^{\rm r}[\Gamma^{\rm r}_{t}|\Gamma^{\rm r}_{0}]},
}
where $p(\Gamma_{0})$ ($p^{\rm r}(\Gamma^{\rm r}_{0})$) and $\mathcal{P}[\Gamma_{t}|\Gamma_{0}]$ ($\mathcal{P}^{\rm r}[\Gamma^{\rm r}_{t}|\Gamma^{\rm r}_{0}]$) are the initial distribution and the conditional probability of the forward (reference) process, respectively. The reference process will be chosen to be the backward or backward-dual process depending on the system under consideration. Here the dual process is defined to be the process whose steady state is the same as the original one except that the direction of the stationary current is opposite to the original one \cite{HS07,Sei12,EB10}.   
We first consider the conditional ensemble average of $e^{-\sigma}$ on condition that the measurement outcome is $\{m^{N}\}$:
\eqn{
 \langle e^{-\sigma}|\{m^{N}\}\rangle&=&\frac{1}{P\bigl ( \{m^{N}\}\bigr )}\int_{\prod_{k=1}^{N}X_{m_{k}}}\mathcal{D}\Gamma  \mathcal{P}[\Gamma|\Lambda_{\{m^{N}\}}]e^{-\sigma} \nonumber\\
&=&\frac{1}{P\bigl ( \{m^{N}\}\bigr )}\int_{\prod_{k=1}^{N}X^{\rm r}_{m_{k}}}\mathcal{D}\Gamma^{\rm r} \mathcal{P}^{\rm r}[\Gamma^{\rm r}|\Lambda^{\rm r}_{\{m^{N}\}}] \nonumber \\
&=&\frac{\prod_{k=1}^{N}\rho^{\rm r}\bigl(m^{\rm r}_{k}|\{m^{k}\}^{\rm r},\Lambda^{\rm r}(\{m^{N}\})\bigr)}{P\bigl ( \{m^{N}\}\bigr )} \nonumber \\ 
&=&\frac{1-\lambda_{{\rm S},\{m^{N}\}}}{P\bigl ( \{m^{N}\}\bigr )}, 
}
where $\prod_{k=1}^{N}X_{m_{k}}$ indicates that the integration is taken over the paths that go through the region $X_{m_{k}}$ at $t=t_{k}$,  $\rho^{\rm r}\bigl(m^{\rm r}_{k}\big|\{m^{k}\}^{\rm r},\Lambda^{\rm r}(\{m^{N}\})\bigr)$ is the conditional probability of obtaining outcome $m^{\rm r}_{k}$ in the reference process on condition that we performed the reference protocol $\Lambda^{\rm r}\bigl(\{m^{N}\}\bigr)$ and obtained the outcomes $\{m^{k}\}^{\rm r}$ at earlier times. Here $\lambda_{{\rm S},\{m^{N}\}}$ is the weight of the singular part of the reference measure associated with a realization $\{m^{N}\}$. This term quantifies the weight of those paths which have a vanishing probability in the forward process but have a non-zero probability in the backward process.  We define the unavailable information in the same manner as in Eq. (\ref{unavb}) and the integral fluctuation theorem is obtained by
\eqn{
\langle e^{-\sigma-I+I_{\rm{u}}}\rangle&=&\Biggl\langle e^{-\sigma}\prod_{k=1}^{N}\frac{p\bigl(m_{k}|\{m^{k-1}\}\bigr)}{\rho^{\rm r}\bigl(m^{\rm r}_{k}|\{m^{k}\}^{\rm r},\Lambda^{\rm r}(\{m^{N}\})\bigr)}\Biggr\rangle \nonumber \\
&=&\sum_{\{m^{N}\}}P\bigl(\{m^{N}\}\bigr)\langle e^{-\sigma}|\{m^{N}\}\rangle\frac{P\bigl(\{m^{N}\}\bigr)}{1-\lambda_{{\rm S},\{m^{N}\}}} \nonumber \\
&=&1. \label{Master}
}

Our main result can be given by choosing the time-reversed process as the reference process and the equilibrium distribution associated with the corresponding fixed external parameters as the initial probability distribution of the time-reversed process. We also assume that the initial probability distribution in the forward process is the equilibrium distribution associated with the initial value of the external parameters. In this case, $I_{\rm u}$ takes the form of Eq. (\ref{unav}) and $\sigma$ is proportional to the dissipated work:
\eqn{
\sigma=\beta(W-\Delta F)=\beta W_{\rm d}. 
}
Then, Eq. (\ref{Master}) is rewritten as
\eqn{
\langle e^{-\beta W_{\rm d}-I+I_{\rm{u}}}\rangle=1,\label{FT2}
}
where we assume  the detailed fluctuation theorem \cite{Cro98,Jar00}, $\mathcal{P}[\Gamma_{t}|\Gamma_{0}]/\mathcal{P}^{r}[\Gamma^{\rm r}_{t}|\Gamma^{\rm r}_{0}]=e^{-\beta Q}$ with $Q$ being a heat flowing from a bath to the system. Equation (\ref{FT2}) was first obtained  for the system without feedback control \cite{Jar97Apr} and then generalized to the system under feedback control \cite{SU10}. Note that the derivation  in Ref. \cite{SU10} is not applicable to the case of error-free measurements. 
Appling Jensen's inequality to Eq. (\ref{FT2}), we obtain the following inequality:
\eqn{\label{sec}
-\langle W_{\rm d} \rangle \leq k_{\rm B}T(\langle I \rangle -\langle I_{\rm{u}} \rangle),
} 
which means that the new term, unavailable information, reduces the achievable bound of extractable work and leads to a more stringent inequality than the inequality (\ref{ineq1}). Note that since the feedback protocol designed for work extraction is supposed to extract work maximally when the measurements are perfect, the RHS of (\ref{sec})  gives an upper bound even if there exist errors in the measurements.
The equality in (\ref{sec}) is achieved if and only if $W_{\rm d}$ does not fluctuate for each realization of the feedback process. Therefore, the RHS of (\ref{sec}) gives an achievable upper bound of extractable work for a given feedback protocol. We shall explicitly illustrate this by examples  in Sec. \ref{ex}. 

 We emphasize that the known equalities in Refs. \cite{SU10,AS12,SJ12} break down under  error-free measurements because the conventional approaches cannot deal with the case in which the probability of forward paths vanishes due to error-free measurements.  We have obtained the equality (\ref{Master}) which is applicable to error-free measurements by precisely quantifying the contribution of singular paths which leads to the information loss in the feedback process.
   
   Equation (\ref{Master}) leads to various integral fluctuation theorems by an appropriate choice of the reference process. Here we discuss two well-known cases.

\subsection*{Seifert's form}
We choose a reference process as the time-reversed process of the forward process. Let $p(\Gamma_{0})$ be an arbitrary initial distribution and $p^{\rm r}(\Gamma^{\rm r}_{0})$ be the final distribution of the forward process. In this case, $\sigma$ is equivalent to the total entropy production: 
\eqn{
\sigma=\Delta s_{\rm tot}\equiv\Delta s_{\rm sys}+\Delta s_{\rm m},
}
where $\Delta s_{\rm sys}\equiv -\ln p(\Gamma_{\tau};\tau)+\ln p(\Gamma_{0};0)$ and $\Delta s_{\rm m}\equiv -\beta Q$ is the entropy production of the system and the bath, respectively.

We can now rewrite Eq. (\ref{Master}) and obtain the following integral fluctuation theorem, which was first derived for systems without feedback control by Seifert \cite{Sei05} and later generalized to systems under feedback control by Lahiri {\it et al.} \cite{SJ12}:
\eqn{
\langle e^{-\Delta s_{\rm tot}-I+I_{\rm{u}}}\rangle =1. \label{sei}
}

\subsection*{Hatano-Sasa's form}
We next discuss a situation in which the stationary distribution associated with fixed parameters is not an equilibrium one. The reference dynamics should then be chosen to be a time-reversed and dual one \cite{HS07,Sei12,EB10}, and let the initial distribution be a nonequilibirum steady state associated with the corresponding parameters. In this case, $\sigma$ is expressed as
\eqn{
\sigma=\Delta \phi-\beta Q_{\rm{ex}},
}
where $\phi(\lambda)\equiv-\ln p_{\rm{st}}(\lambda)$ and $Q_{\rm{ex}}$ is an excess heat flowing from a bath to the system. Then, Eq. (\ref{Master}) is rewritten in the form of the following integral fluctuation theorem, which was first derived for systems without feedback control by Hatano and Sasa \cite{HS01} and later generalized to systems under feedback control in Refs. \cite{AS12,SJ12}:
\eqn{
\langle e^{-\Delta\phi+\beta Q_{\rm{ex}}-I+I_{\rm{u}}}\rangle =1. \label{HT}
} 
\section{Examples \label{ex}}
To get physical insights into the above results,  we offer simple examples, which achieve the equality in (\ref{sec}) and give a physical interpretation of unavailable information.
\subsection{Brownian particle in a harmonic trap \label{brown}}
\begin{figure}[t]
\includegraphics[scale=0.5]{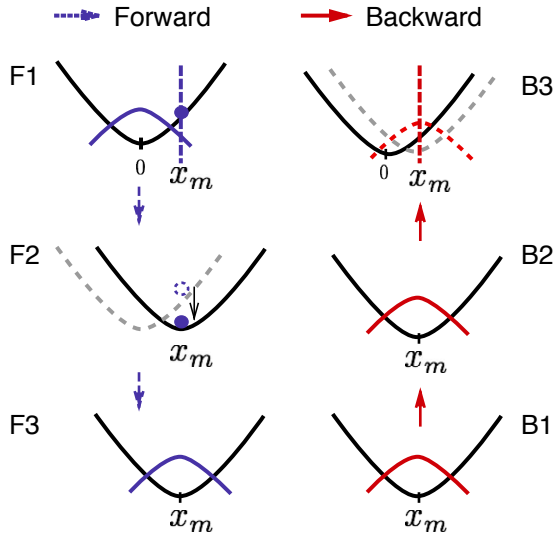}
\caption{\label{figB} Brownian particle in a harmonic potential with the forward (downward dashed arrows) or backward (upward solid arrows) protocol. First, the position of the particle is measured without error (F1). Then, according to the outcome, the potential is instantaneously shifted and work is extracted from the difference in the potential energy (indicated by the downward solid  arrow) (F2). Finally, the system becomes equilibrated with the heat bath (F3). The associated backward process (upward solid arrows) starts from the equilibrium distribution for the final configuration of the forward process (B1, B2). Then the potential is instantaneously shifted back to the initial position (B3).  Some amount of information gets loss and becomes unavailable in the ensuing relaxation process (from F2 to F3).}
\end{figure}
We consider a Brownian particle in a harmonic potential \cite{AS11} whose center $\lambda$ is manipulated to extract work according to the outcome $x_{\rm m}$ of the position measurement (Fig. \ref{figB}).
The system is initially prepared to be at thermal equilibrium with a heat bath with inverse temperature $\beta$, and the error-free position  measurement is performed (F1). Then the potential is instantaneously shifted from $\lambda_{\rm i}=0$ to $\lambda_{\rm f}=x_{\rm m}$ and work is extracted from the difference in the potential energy (F2). Finally, the relaxation occurs and the system becomes equilibrated with the heat bath (F3). From F2 to F3, some amount of information about the position of the particle is lost and becomes unavailable. 

The information acquired by the measurement can be calculated by averaging  Eq. (\ref{Sh}) over all possible realizations of the protocol:
\eqn{
\langle I\rangle&=&-\int_{-\infty}^{\infty}dx_{\rm m} p_{\rm eq}(x_{\rm m})\ln p_{\rm eq}(x_{\rm m})\nonumber \\
&=&\frac{1}{2}-\ln\sqrt{\frac{\beta k}{2\pi}},
}
where $k$ is the spring constant and $p_{\rm eq}(x)$ is the equilibrium distribution: $p_{\rm eq}(x)=\sqrt{\beta k/2\pi}\exp(-\beta k x^{2}/2).$
Here, note that we ignore a divergent term $\ln (dx)$ arising from the infinitely precise measurements since we only consider the difference (the RHS of (\ref{sec})) rather than the absolute  amount of information. 

Let us next calculate the unavailable information. To this end, we consider the backward process: the system is initially in equilibrium and the center of the potential is located at $\lambda^{\rm R}_{\rm i}=x_{\rm m}$ (B1, B2). Then the potential is instantaneously shifted to $\lambda^{\rm R}_{\rm f}=0$ (B3). At this stage, the probability of the particle being found at $x_{\rm m}$ is equal to $p_{\rm eq}(0)$ and therefore, the unavailable information is given by averaging Eq. (\ref{unav}) over all possible realizations of the protocol:
\eqn{
\langle I_{\rm{u}}\rangle&=&-\int_{-\infty}^{\infty}dx_{\rm m} p_{\rm eq}(x_{\rm m})\ln p_{\rm eq}(0)=-\ln\sqrt{\frac{\beta k}{2\pi}}. \nonumber
}
The achievable upper bound of extractable work is now obtained from the RHS of (\ref{sec}) as
\eqn{\label{upboundB}
k_{\rm B}T\bigl(\langle I\rangle -\langle I_{\rm{u}}\rangle \bigr)=\frac{k_{\rm B}T}{2}.
}
On the other hand, the extracted work is 
\eqn{
-\langle W\rangle=\int_{-\infty}^{\infty}dx p_{\rm eq}(x)\frac{kx^{2}}{2}=\frac{k_{\rm B}T}{2}.
} 
Since $\langle \Delta F\rangle=0$, the equality in (\ref{sec}) is achieved in this example. This is because  the process is performed instantaneously and therefore $W$ does not fluctuate for each realization of the feedback protocol. If the measurements are performed with errors, extracted work is less than the value in Eq. (\ref{upboundB}) \cite{AS11}. Hence we conclude that the inequality (\ref{sec}) indeed gives the achievable upper bound of extractable work. Note that since $I$ diverges in infinitely precise measurements, the inequality (\ref{ineq1}) does not give an achievable bound. This is because the inequality (\ref{ineq1}) does not take into account the contribution from the unavailable information. 

We can also  confirm the integral fluctuation theorem (\ref{FT2}):
\eqn{
\langle e^{-\beta W_{\rm d}-I+I_{\rm{u}}} \rangle =\int_{-\infty}^{\infty}dx p_{\rm eq}(x)e^{\frac{\beta k x^{2}}{2}}\frac{p_{\rm eq}(x)}{p_{\rm eq}(0)}=1,\nonumber
}
where we have substituted the dissipated work $W_{\rm d}=-kx^{2}/2$, the acquired information $I=-\ln p_{\rm eq}(x)$, and the unavailable information $I_{\rm u}=-\ln p_{\rm eq}(0)$.  
\subsection{Classical Szilard Engine}
We discuss a generalized classical Szilard engine \cite{SU12,Szi29}, as schematically illustrated in Fig. \ref{figS}. In the figure, the forward process is indicated by dashed downward arrows and the backward process is indicated by solid upward ones. (a) A single-particle classical gas is confined in a box and equilibrates with a surrounding heat bath at inverse temperature $\beta$. (b) A barrier is inserted at position $l_{\rm i}$ and a measurement is performed to determine which (left (L) or right (R)) partition the particle is in. The information acquired by the measurement can be evaluated as
\eqn{\label{acin}
\langle I\rangle=-l_{\rm i}\ln l_{\rm i}-(1-l_{\rm i})\ln (1-l_{\rm i}).
} Note that the length of the box is set to $L=1$ for simplicity.
(c) We quasi-statistically shift the position of the barrier to $l_{\rm L}$ or $l_{\rm R}$ depending on whether the outcome of the measurement is L or R. (d) Finally, the barrier is removed and the system gets equilibrated with the heat bath. During the process of removal of the barrier and the subsequent thermal relaxation (from (c) to (d)), a certain amount of information has been lost and become unavailable for use in extracting work.  Note that no work is needed to insert or remove the barrier since we consider the classical case.

To evaluate the unavailable information, let us consider the backward process as indicated by solid upward arrows in Fig. \ref{figS}. The system is initially prepared to be at thermal equilibrium in (d) and a barrier is inserted at  position $l_{\rm L}$ or $l_{\rm R}$ according to the measurement outcome in the forward process in (c). At this stage, the singular paths arises; there are no counterparts in the forward process for paths such that a particle resides in the right (left) side, although the barrier is inserted at $l_{\rm L}$ ($l_{\rm R}$)  as indicated by the leftmost (rightmost) path in Fig. \ref{figS}. Then the barrier is moved to $l_{\rm i}$ in (b) and finally removed in (a). 

The probability weight of the singular paths associated with each realization is given by:
\eqn{
\lambda_{\rm S,L}=1-l_{\rm L} \nonumber, \:\:  \lambda_{\rm S,R}=l_{\rm R}\nonumber.
}
This is because the probability of obtaining the outcome R (L) in the backward process on condition that the outcome is L (R) in the forward process is equal to  $1-l_{\rm{L}}$ ($l_{\rm{R}}$). 
We therefore obtain the unavailable information by averaging Eq. (\ref{unavsin}) over all possible realizations of the protocol:
\eqn{\label{unavin}
\langle I_{\rm{u}}\rangle=-l_{\rm i}\ln l_{\rm L}-(1-l_{\rm i})\ln (1-l_{\rm R}). 
}
On the other hand, the extracted work is given by
\eqn{\label{work}
-\langle W\rangle=-k_{\rm B}T\biggl[ l_{\rm i}\ln \frac{l_{\rm i}}{l_{\rm L}}+(1-l_{\rm i})\ln \biggl(\frac{1-l_{\rm i}}{1-l_{\rm R}}\biggr) \biggr].
}
From Eqs. (\ref{acin}), (\ref{unavin}), and (\ref{work}), we obtain
\eqn{
-\langle W\rangle=k_{\rm B}T\bigl(\langle I\rangle-\langle I_{\rm{u}}\rangle\bigr) \leq k_{\rm B}T\langle I\rangle.
}
This means that the inequality (\ref{sec}) gives an achievable bound of extractable work while the inequality (\ref{ineq1}) does not because some amount of information becomes unavailable in free expansion which ensues after the removal of the barrier. In this example, the equality in (\ref{sec}) is achieved because the process is performed quasi-statistically and therefore $W$ does not fluctuate for each realization. We can also confirm the integral fluctuation theorem (\ref{FT2}):
\eqn{
\langle e^{-\beta W_{d}-I+I_{\rm{u}}} \rangle =l_{\rm i}\frac{l_{\rm L}}{l_{\rm i}}\frac{l_{\rm i}}{l_{\rm L}}+(1-l_{\rm i})\frac{1-l_{\rm R}}{1-l_{\rm i}}\frac{1-l_{\rm i}}{1-l_{\rm R}}=1.\nonumber
}

 An optimal protocol, {\it i.e.,} the protocol that maximizes the extractable work, can be designed by choosing $l_{\rm i}=1/2$, $l_{\rm L}=1$, and $l_{\rm R}=0$. In this case, there are no singular paths and the unavailable information is zero. Therefore, the inequality (\ref{ineq1}) can be achieved in a single-particle Szilard engine. However, this is not the case for a multi-particle Szilard engine as discussed in the next subsection. 
\subsection{Quantum Szilard Engine \label{qsze}}
We next discuss a quantum Szilard engine \cite{Kim11} in which the work needed to insert or remove the barrier is taken into account by evaluating the partition function in a quantum statistical mechanical manner. The protocol is a straightforward  generalization of the classical Szilard engine and defined as follows. First, an $N$-particle non-interacting  quantum gas is confined in a trapping potential and equilibrates with a heat bath. A barrier is inserted at position $l_{\rm i}$ and the number of particles in the left box, $m$, is measured.  The barrier is then moved to $l_{m}$ according to the measurement outcome $m$.  Finally, the barrier is removed and the system returns to the initial state. All these  processes are performed isothermally and the coherence between the system and a bath is assumed to be completely destroyed every moment. 

 The acquired information is calculated by evaluating the partition function of the system. Let $Z_{m}(l_{\rm i})$ be the partition function of the system when the barrier is inserted at $l_{\rm i}$ and the number of particles in the left (right) side is $m$ ($N-m$). The probability of finding outcome $m$ is 
 \eqn{
p(m)=\frac{Z_{m}(l_{\rm i})}{Z(l_{\rm i})},
 }
 where $Z(l_{\rm i})\equiv\sum_{m=1}^{N}Z_{m}(l_{\rm i})$ is the partition function of the system before the measurement. The acquired information is now given by
 \eqn{\label{I}
\langle I\rangle=-\sum_{m=1}^{N}p(m)\ln p(m).
 }
 
 Let us next consider the contribution from the unavailable information. The backward process is defined in the same way as in the single-particle Szilard engine (see Fig. \ref{figS}). We first calculate the probability weight of the singular paths $\lambda_{{\rm S},m}$ corresponding to the protocol $\Lambda(m)$ which is the protocol associated with the measurement outcome $m$:
 \eqn{
 \lambda_{{\rm S},m}=1-\frac{Z_{m}(l_{m})}{Z(l_{m})}.
}
This is because the probability of not finding the outcome $m$ in the backward process $\Lambda^{\rm R}(m)$ is equal to $1-Z_{m}(l_{m})/Z(l_{m})$.
Therefore, the unavailable information is obtained by averaging Eq. (\ref{unavsin}) over all possible realizations of the protocol:
\eqn{\label{Iu}
\langle I_{\rm{u}} \rangle=-\sum_{m=1}^{N}p(m)\ln\frac{Z_{m}(l_{m})}{Z(l_{m})}.
}Physically, the unavailable information arises because some amount of information is lost in free expansion upon the removal of the barrier.

On the other hand, the total extracted work can be obtained by taking into account the work needed to insert or remove the barrier. The result is \cite{Kim11}
\eqn{\label{W}
-\langle W\rangle=-k_{\rm B}T\sum_{m=1}^{N}p(m)\ln \Bigl(\frac{p(m)Z(l_{m})}{Z_{m}(l_{m})}\Bigr).
}
From Eqs. (\ref{I}), (\ref{Iu}), and (\ref{W}), we can confirm that the inequality (\ref{sec}) is saturated:
\eqn{\label{sat}
-\langle W\rangle=k_{\rm B}T\bigl(\langle I\rangle -\langle I_{\rm{u}}\rangle\bigr).
}
This is because all processes are performed quasi-statistically and $W$ does not fluctuate from one realization to another. Note that there exists a non-zero contribution from $\lambda_{{\rm S},m}$ for any choices of parameters $l_{{\rm i},m}$ in a multi-particle case and hence, some amount of information inevitably becomes unavailable, {\it i.e.,} $\langle I_{\rm{u}}\rangle>0$. Therefore, the inequality (\ref{ineq1}) does not give an achievable bound for a multi-particle Szilard engine, whereas our inequality (\ref{sec}) gives an achievable upper bound of extractable work for any particle number.
We can also confirm the integral fluctuation theorem:
\eqn{
\langle e^{-\beta W_{d}-I+I_{\rm{u}}}\rangle=\sum_{m=1}^{N}p(m)\frac{Z_{m}(l_{m})}{p(m)Z(l_{m})}p(m)\frac{Z(l_{m})}{Z_{m}(l_{m})}=1, \nonumber
}
where we have substituted the dissipated work $W_{\rm d}=k_{\rm B}T\ln \bigl(p(m)Z(l_{m})/Z_{m}(l_{m})\bigr)$, the acquired information $I=-\ln p(m)$, and the unavailable information $I_{\rm u}=-\ln (Z_{m}(l_{m})/Z(l_{m}))$.

We next briefly discuss the optimal protocol which is the protocol that maximizes an extractable work. To this end, we note that from Eqs. (\ref{Iu}) and (\ref{sat}) the optimum choice of $l_{m}$ can be obtained by minimizing the unavailable information, which amounts to maximizing $Z_{m}(l_{m})/Z(l_{m})$ with respect to $l_{m}$. On the basis of this observation, we can now clarify the controversy conducted in Refs. \cite{POV13,Kim13}. In the originally proposed model \cite{Kim11}, Kim {\it et al.} have chosen the value of $l_{m}$ such that $Z_{m}(l_{m})$ takes the maximum value. However, this choice is not optimal in general as pointed out by Plesch {\it et al.} \cite{POV13}. This is not because the expression (\ref{W}) is incorrect \cite{POV13} but because $Z(l)$  depends on $l$, and hence maximizing $Z_{m}(l)$ does not necessarily lead to minimizing unavailable information, {\it i.e.}, maximizing $Z_{m}(l_{m})/Z(l_{m})$. As a result, the protocol proposed by Kim {\it et al.} \cite{Kim11} is not optimal.

Furthermore, we can also consider maximizing an extractable work by adjusting the position of the initial insertion $l_{\rm i}$. To do so, we note that Eq. (\ref{W}) can be converted to 
\eqn{\label{opt}
-\langle W\rangle=k_{\rm B}T\bigl(\ln \gamma -D(p(m)||\hat{\rho}^{\rm R}(m))\bigr),
}
where $\gamma=\sum_{m=1}^{N}Z_{m}(l_{m})/Z(l_{m})$, $\hat{\rho}^{\rm R}(m)=Z_{m}(l_{m})/(\gamma Z(l_{m}))$, and $D$ is the Kullback-Leibler divergence \cite{Cov06}. Note that $\gamma$ is equal to an efficacy parameter of feedback control \cite{SU10,SU12} and $\hat{\rho}^{\rm R}(m)$ is normalized to unity: $\sum_{m=1}^{N}\hat{\rho}^{\rm R}(m)=1$. From Eq. (\ref{opt}) and the fact that $p(m)$ only depends on $l_{\rm i}$, we can maximize an extractable work by minimizing $D(p(m)||\hat{\rho}^{\rm R}(m))$ with respect to $l_{\rm i}$. The above considerations lead to guiding principles of designing the optimal protocol as discussed in the next subsection from a general perspective.
\subsection{Implications}
We here discuss some implications from previous examples. First, we note that the contribution of the unavailable information also exists in recent experimental realizations using a Brownian particle \cite{LBK08,TSUe10}. In those experiments, the origin of unavailable information is essentially the same as in the first example in Sec. \ref{brown}: some amount of information is lost and becomes unavailable through the relaxation.

Second, we can obtain guiding principles of designing the optimal protocol in the case of the feedback control with a single measurement by generalizing the discussion in Sec.\ref{qsze}. This consists of the following two steps:
\begin{enumerate}
\item Design the protocol {\it after} the measurement such that the unavailable information is minimized.
\item Design the protocol {\it before} the measurement such that $D(P(m)||\hat{\rho}^{\rm R}(m))$ is minimized,
\end{enumerate}
where $m$ labels the measurement outcome. From Eq. (\ref{unavsin}), the first step amounts to designing each protocol $\Lambda(m)$  such that the singular part $\lambda_{{\rm S}, m}$ is minimized. The second step is inferred from the following inequality 
\eqn{
-\langle W_{\rm d}\rangle\leq k_{\rm B}T\bigl(\ln \gamma -D(P(m)||\hat{\rho}^{\rm R}(m))\bigr),
}
which is derived from the inequality (\ref{sec}) in the same manner  as in Eq. (\ref{opt}). This can be  interpreted as follows: an extractable work can be maximized by assigning a higher probability of realization to the protocol in which the unavailable information is less.
We can illustrate the above guiding principles by considering the quantum Szilard engine; the first step corresponds to choosing $l_{m}$ such that $Z_{m}(l_{m})/Z(l_{m})$ takes the maximum value and the second step corresponds to choosing the initial insertion position $l_{\rm i}$ so as to minimize $D(p(m)||\hat{\rho}^{\rm R}(m))$. 

Note that the guiding principles proposed here are inapplicable to the system subject to repeated measurements. This is because the probability distribution of the measurement outcomes is independent of the protocol after the first measurement only if the measurement is performed once. Finding the guiding principles for the feedback control subject to repeated measurements remains an interesting open question.
\section{Conclusion \label{con}}  
We have obtained a general achievable upper bound of extractable work for a given feedback protocol. The inequality (\ref{sec}) is a consequence of the nonequilibrium equality under error-free measurements (\ref{FT2}). The new term, $I_{\rm{u}}$, quantifies the amount of information which becomes unavailable in the feedback process. We relate the unavailable information (\ref{unavsin}) to the singular part of the reference measure. The variants of the integral fluctuation theorem (\ref{sei}) and (\ref{HT}) are also obtained. Although the original derivations in Refs. \cite{SU10,AS12,SJ12} cannot be applied to error-free measurements, we point out that our results remain valid under such a situation.
 
 As an illustration, a Brownian particle in a harmonic potential, and classical and quantum Szilard engines are discussed and shown to achieve the equality in (\ref{sec}). In the former example, the contribution from the unavailable information needs to be taken into account since some amount of information is lost and becomes unavailable through the relaxation. In the latter examples, some amount of information becomes lost in free expansions after the removal of a barrier. We also show that the inequality (\ref{ineq1}) does not give an achievable bound of extractable work in a multi-particle Szilard engine. The guiding principles of designing the optimal protocol are proposed.
 
An interesting future direction is to investigate our result experimentally or numerically in genuine nonequilibrium systems. We give the discussion about experimental implementation of an error-free measurement in Appendix. Other interesting direction is to consider similar thermodynamic relations in autonomous Maxwell's demon systems in which systems evolve without external agencies.

\begin{acknowledgments}
This work was supported by
KAKENHI Grant No. 26287088 from the Japan Society for the Promotion of Science, 
and a Grant-in-Aid for Scientific Research on Innovation Areas ``Topological Quantum Phenomena" (KAKENHI Grant No. 22103005),
and the Photon Frontier Network Program from MEXT of Japan.
K. F. acknowledges support from JSPS (Grant No. 254105). Y. M. was supported by Japan Society for the Promotion of Science through Program for Leading Graduate Schools (MERIT). Y. A. thanks S. Toyabe for a useful discussion.
\end{acknowledgments}

\appendix*
\section{}
 Here we discuss experimental implementation of an error-free measurement. The crucial point is that if the contribution of non-zero measurement error is masked by experimental statistical error, we can regard a measurement as an error-free one within the experimental accuracy. To elucidate this point, we here consider a Brownian particle in a harmonic potential as a concrete example.
 
 Let us assume that a Brownian particle is initially at thermal equilibrium with a heat bath with inverse temperature $\beta$ and the center of the harmonic trap is located at $\lambda=0$. We measure which side of the trap the particle resides in, {\it i.e.}, the measurement outcome is left (right) if the particle is founded in $(-\infty, 0]$ ($[0,\infty)$). Then, according to the measurement outcome L or R, we instantaneously shift the center of the potential from $0$ to $-\Delta<0$ or $\Delta>0$  and extract work from the difference in the potential energy.
 
    \begin{figure}[b]
\includegraphics[scale=0.87]{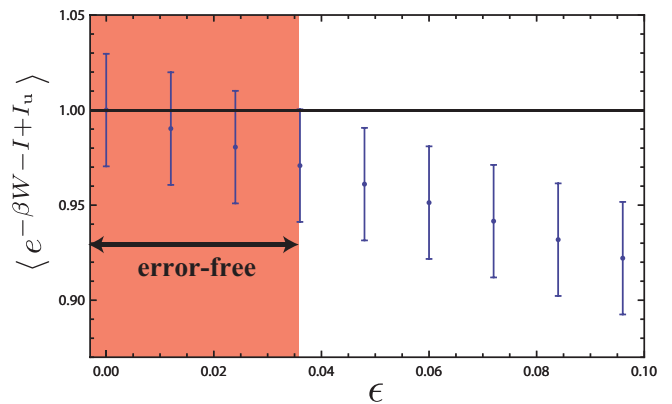}
\caption{\label{figstat-error} (Color online) The criterion of experimental implementation of an error-free measurement in the model of a Brownian particle in a harmonic potential. The LHS of Eq. (\ref{FT2}) (vertical line) is plotted against the measurement error probability $\epsilon$ (horizontal line) for $\sqrt{\beta k}\Delta=1.0$ and $N=1000$. The shaded region implies that the deviation of the value of  $\langle e^{-\beta W-I+I_{\rm u}}\rangle$  from 1 is masked by the statistical error in that region so that the measurement can be regarded as error-free.}
\end{figure}
 
 If the measurement is performed perfectly, we can check that the integral fluctuation theorem (\ref{FT2}) holds in the same manner as in Sec. \ref{brown}. However,  practically there is always non-zero measurement error probability in experimental situations and this should be taken into account if we consider the experimental verification of Eq. (\ref{FT2}).
 
 To do so, here we simply model an error probability $\epsilon$ as follows: if the true position of the particle is in $(-\infty, 0]$ ($[0,\infty)$), then the measurement outcome is L (R) with probability $1-\epsilon$ and R (L) with probability $\epsilon$. Under this assumption, the LHS of Eq. (12) now deviates from 1 and we obtain
 \eqn{
 \langle e^{-\beta W-I+I_{\rm u}}\rangle=1+\delta(\epsilon),
 }
 \eqn{
 \delta(\epsilon)=\epsilon\Bigl(\frac{1}{P(\Delta)}-2\Bigr),
 }
\eqn{
P(\Delta)=\int_{-\infty}^{\Delta}p_{\rm eq}(x) dx,
}
where  $p_{\rm eq}(x)$ is the equilibrium distribution: $p_{\rm eq}(x)=\sqrt{\beta k/2\pi}\exp(-\beta k x^{2}/2)$ and $k$ is the spring constant. 

On the other hand, it is also true that the trial number $N$ of experiments is limited and there is always non-zero statistical error $\sigma(N)$. We estimate $\sigma(N)$ by the standard deviation:
\eqn{
\sigma(N)&=&\sqrt{\frac{1}{N}\Bigl(\langle(e^{-\beta W-I+I_{\rm u}})^{2}\rangle-\langle e^{-\beta W-I+I_{\rm u}}\rangle^{2}\Bigr)} \nonumber \\
&=&\sqrt{\frac{1}{N}\Bigl(\frac{e^{\beta k\Delta^{2}}P(2\Delta)}{2P(\Delta)^{2}}-1\Bigr)},
}
where we neglect the contribution of the measurement error $\epsilon$.

The crucial point is that if the measurement error $\epsilon$ is sufficiently small so that the deviation $\delta(\epsilon)$ is masked by the statistical error $\sigma(N)$, $\delta(\epsilon)<\sigma(N)$, then the measurement can be regarded as error-free for the purpose of the experimental verification of Eq. (\ref{FT2}).  To quantify this augment, we plot $1+\delta(\epsilon)\pm\sigma(N)$ in Fig. \ref{figstat-error}. As shown in Fig. \ref{figstat-error}, experimentalists need to determine which side (L or R) of the potential the particle resides in with an error less than about 4\% to implement an error-free measurement. This accuracy can be achieved by current experimental technologies \cite{TY}.  Although this criterion of an error-free measurement depends on the details of each experimental setup, one can follow a similar discussion and estimate the accuracy required for the experimental verification of our thermodynamic relation (\ref{FT2}).
\bibliography{reference}

\begin{thebibliography}{47}%
\makeatletter
\providecommand \@ifxundefined [1]{%
 \@ifx{#1\undefined}
}%
\providecommand \@ifnum [1]{%
 \ifnum #1\expandafter \@firstoftwo
 \else \expandafter \@secondoftwo
 \fi
}%
\providecommand \@ifx [1]{%
 \ifx #1\expandafter \@firstoftwo
 \else \expandafter \@secondoftwo
 \fi
}%
\providecommand \natexlab [1]{#1}%
\providecommand \enquote  [1]{``#1''}%
\providecommand \bibnamefont  [1]{#1}%
\providecommand \bibfnamefont [1]{#1}%
\providecommand \citenamefont [1]{#1}%
\providecommand \href@noop [0]{\@secondoftwo}%
\providecommand \href [0]{\begingroup \@sanitize@url \@href}%
\providecommand \@href[1]{\@@startlink{#1}\@@href}%
\providecommand \@@href[1]{\endgroup#1\@@endlink}%
\providecommand \@sanitize@url [0]{\catcode `\\12\catcode `\$12\catcode
  `\&12\catcode `\#12\catcode `\^12\catcode `\_12\catcode `\%12\relax}%
\providecommand \@@startlink[1]{}%
\providecommand \@@endlink[0]{}%
\providecommand \url  [0]{\begingroup\@sanitize@url \@url }%
\providecommand \@url [1]{\endgroup\@href {#1}{\urlprefix }}%
\providecommand \urlprefix  [0]{URL }%
\providecommand \Eprint [0]{\href }%
\providecommand \doibase [0]{http://dx.doi.org/}%
\providecommand \selectlanguage [0]{\@gobble}%
\providecommand \bibinfo  [0]{\@secondoftwo}%
\providecommand \bibfield  [0]{\@secondoftwo}%
\providecommand \translation [1]{[#1]}%
\providecommand \BibitemOpen [0]{}%
\providecommand \bibitemStop [0]{}%
\providecommand \bibitemNoStop [0]{.\EOS\space}%
\providecommand \EOS [0]{\spacefactor3000\relax}%
\providecommand \BibitemShut  [1]{\csname bibitem#1\endcsname}%
\let\auto@bib@innerbib\@empty
\bibitem [{\citenamefont {Evans}\ \emph {et~al.}(1993)\citenamefont {Evans},
  \citenamefont {Cohen},\ and\ \citenamefont {Morriss}}]{ECM93}%
  \BibitemOpen
  \bibfield  {author} {\bibinfo {author} {\bibfnamefont {D.~J.}\ \bibnamefont
  {Evans}}, \bibinfo {author} {\bibfnamefont {E.~G.~D.}\ \bibnamefont {Cohen}},
  \ and\ \bibinfo {author} {\bibfnamefont {G.~P.}\ \bibnamefont {Morriss}},\
  }\href@noop {} {\bibfield  {journal} {\bibinfo  {journal} {Phys. Rev. Lett.}\
  }\textbf {\bibinfo {volume} {71}},\ \bibinfo {pages} {2401} (\bibinfo {year}
  {1993})}\BibitemShut {NoStop}%
\bibitem [{\citenamefont {Gallavotti}\ and\ \citenamefont
  {Cohen}(1995)}]{GC95}%
  \BibitemOpen
  \bibfield  {author} {\bibinfo {author} {\bibfnamefont {G.}~\bibnamefont
  {Gallavotti}}\ and\ \bibinfo {author} {\bibfnamefont {E.~G.~D.}\ \bibnamefont
  {Cohen}},\ }\href@noop {} {\bibfield  {journal} {\bibinfo  {journal} {Phys.
  Rev. Lett.}\ }\textbf {\bibinfo {volume} {74}},\ \bibinfo {pages} {2694}
  (\bibinfo {year} {1995})}\BibitemShut {NoStop}%
\bibitem [{\citenamefont {Kurchan}(1998)}]{Kur98}%
  \BibitemOpen
  \bibfield  {author} {\bibinfo {author} {\bibfnamefont {J.}~\bibnamefont
  {Kurchan}},\ }\href@noop {} {\bibfield  {journal} {\bibinfo  {journal} {J.
  Phys. A: Math. Gen.}\ }\textbf {\bibinfo {volume} {31}},\ \bibinfo {pages}
  {3719} (\bibinfo {year} {1998})}\BibitemShut {NoStop}%
\bibitem [{\citenamefont {Lebowitz}\ and\ \citenamefont {Spohn}(1999)}]{LS99}%
  \BibitemOpen
  \bibfield  {author} {\bibinfo {author} {\bibfnamefont {J.~L.}\ \bibnamefont
  {Lebowitz}}\ and\ \bibinfo {author} {\bibfnamefont {H.}~\bibnamefont
  {Spohn}},\ }\href@noop {} {\bibfield  {journal} {\bibinfo  {journal} {J.
  Stat. Phys.}\ }\textbf {\bibinfo {volume} {95}},\ \bibinfo {pages} {333}
  (\bibinfo {year} {1999})}\BibitemShut {NoStop}%
\bibitem [{\citenamefont {Maes}(1999)}]{Mae99}%
  \BibitemOpen
  \bibfield  {author} {\bibinfo {author} {\bibfnamefont {C.}~\bibnamefont
  {Maes}},\ }\href@noop {} {\bibfield  {journal} {\bibinfo  {journal} {J. Stat.
  Phys.}\ }\textbf {\bibinfo {volume} {95}},\ \bibinfo {pages} {367} (\bibinfo
  {year} {1999})}\BibitemShut {NoStop}%
\bibitem [{\citenamefont {Jarzynski}(1997)}]{Jar97Apr}%
  \BibitemOpen
  \bibfield  {author} {\bibinfo {author} {\bibfnamefont {C.}~\bibnamefont
  {Jarzynski}},\ }\href@noop {} {\bibfield  {journal} {\bibinfo  {journal}
  {Phys. Rev. Lett.}\ }\textbf {\bibinfo {volume} {78}},\ \bibinfo {pages}
  {2690} (\bibinfo {year} {1997})}\BibitemShut {NoStop}%
\bibitem [{\citenamefont {Jarzynski}(2000)}]{Jar00}%
  \BibitemOpen
  \bibfield  {author} {\bibinfo {author} {\bibfnamefont {C.}~\bibnamefont
  {Jarzynski}},\ }\href@noop {} {\bibfield  {journal} {\bibinfo  {journal} {J.
  Stat. Phys.}\ }\textbf {\bibinfo {volume} {98}},\ \bibinfo {pages} {77}
  (\bibinfo {year} {2000})}\BibitemShut {NoStop}%
\bibitem [{\citenamefont {Crooks}(1998)}]{Cro98}%
  \BibitemOpen
  \bibfield  {author} {\bibinfo {author} {\bibfnamefont {G.~E.}\ \bibnamefont
  {Crooks}},\ }\href@noop {} {\bibfield  {journal} {\bibinfo  {journal} {J.
  Stat. Phys.}\ }\textbf {\bibinfo {volume} {90}},\ \bibinfo {pages} {1481}
  (\bibinfo {year} {1998})}\BibitemShut {NoStop}%
\bibitem [{\citenamefont {Crooks}(1999)}]{Cro99}%
  \BibitemOpen
  \bibfield  {author} {\bibinfo {author} {\bibfnamefont {G.~E.}\ \bibnamefont
  {Crooks}},\ }\href@noop {} {\bibfield  {journal} {\bibinfo  {journal} {Phys.
  Rev. E}\ }\textbf {\bibinfo {volume} {60}},\ \bibinfo {pages} {2721}
  (\bibinfo {year} {1999})}\BibitemShut {NoStop}%
\bibitem [{\citenamefont {Hatano}\ and\ \citenamefont {Sasa}(2001)}]{HS01}%
  \BibitemOpen
  \bibfield  {author} {\bibinfo {author} {\bibfnamefont {T.}~\bibnamefont
  {Hatano}}\ and\ \bibinfo {author} {\bibfnamefont {S.-i.}\ \bibnamefont
  {Sasa}},\ }\href@noop {} {\bibfield  {journal} {\bibinfo  {journal} {Phys.
  Rev. Lett.}\ }\textbf {\bibinfo {volume} {86}},\ \bibinfo {pages} {3463}
  (\bibinfo {year} {2001})}\BibitemShut {NoStop}%
\bibitem [{\citenamefont {Kawai}\ \emph {et~al.}(2007)\citenamefont {Kawai},
  \citenamefont {Parrondo},\ and\ \citenamefont {{Van den Broeck}}}]{KpvB07}%
  \BibitemOpen
  \bibfield  {author} {\bibinfo {author} {\bibfnamefont {R.}~\bibnamefont
  {Kawai}}, \bibinfo {author} {\bibfnamefont {J.~M.~R.}\ \bibnamefont
  {Parrondo}}, \ and\ \bibinfo {author} {\bibfnamefont {C.}~\bibnamefont {{Van
  den Broeck}}},\ }\href@noop {} {\bibfield  {journal} {\bibinfo  {journal}
  {Phys. Rev. Lett.}\ }\textbf {\bibinfo {volume} {98}},\ \bibinfo {pages}
  {080602} (\bibinfo {year} {2007})}\BibitemShut {NoStop}%
\bibitem [{\citenamefont {Wang}\ \emph {et~al.}(2002)\citenamefont {Wang},
  \citenamefont {Sevick}, \citenamefont {Mittag}, \citenamefont {Searles},\
  and\ \citenamefont {Evans}}]{WSMe02}%
  \BibitemOpen
  \bibfield  {author} {\bibinfo {author} {\bibfnamefont {G.~M.}\ \bibnamefont
  {Wang}}, \bibinfo {author} {\bibfnamefont {E.~M.}\ \bibnamefont {Sevick}},
  \bibinfo {author} {\bibfnamefont {E.}~\bibnamefont {Mittag}}, \bibinfo
  {author} {\bibfnamefont {D.~J.}\ \bibnamefont {Searles}}, \ and\ \bibinfo
  {author} {\bibfnamefont {D.~J.}\ \bibnamefont {Evans}},\ }\href@noop {}
  {\bibfield  {journal} {\bibinfo  {journal} {Phys. Rev. Lett.}\ }\textbf
  {\bibinfo {volume} {89}},\ \bibinfo {pages} {050601} (\bibinfo {year}
  {2002})}\BibitemShut {NoStop}%
\bibitem [{\citenamefont {Liphardt}\ \emph {et~al.}(2002)\citenamefont
  {Liphardt}, \citenamefont {Dumont}, \citenamefont {Smith}, \citenamefont
  {Ignacio~Tinoco},\ and\ \citenamefont {Bustamante}}]{LDSe02}%
  \BibitemOpen
  \bibfield  {author} {\bibinfo {author} {\bibfnamefont {J.}~\bibnamefont
  {Liphardt}}, \bibinfo {author} {\bibfnamefont {S.}~\bibnamefont {Dumont}},
  \bibinfo {author} {\bibfnamefont {S.~B.}\ \bibnamefont {Smith}}, \bibinfo
  {author} {\bibfnamefont {J.}~\bibnamefont {Ignacio~Tinoco}}, \ and\ \bibinfo
  {author} {\bibfnamefont {C.}~\bibnamefont {Bustamante}},\ }\href@noop {}
  {\bibfield  {journal} {\bibinfo  {journal} {Science}\ }\textbf {\bibinfo
  {volume} {296}},\ \bibinfo {pages} {1832} (\bibinfo {year}
  {2002})}\BibitemShut {NoStop}%
\bibitem [{\citenamefont {Douarche}\ \emph {et~al.}(2005)\citenamefont
  {Douarche}, \citenamefont {Ciliberto}, \citenamefont {Petrosyan},\ and\
  \citenamefont {Rabbiosi}}]{DCPR05}%
  \BibitemOpen
  \bibfield  {author} {\bibinfo {author} {\bibfnamefont {F.}~\bibnamefont
  {Douarche}}, \bibinfo {author} {\bibfnamefont {S.}~\bibnamefont {Ciliberto}},
  \bibinfo {author} {\bibfnamefont {A.}~\bibnamefont {Petrosyan}}, \ and\
  \bibinfo {author} {\bibfnamefont {I.}~\bibnamefont {Rabbiosi}},\ }\href@noop
  {} {\bibfield  {journal} {\bibinfo  {journal} {Europhys. Lett.}\ }\textbf
  {\bibinfo {volume} {70}},\ \bibinfo {pages} {593} (\bibinfo {year}
  {2005})}\BibitemShut {NoStop}%
\bibitem [{\citenamefont {Collin}\ \emph {et~al.}(2005)\citenamefont {Collin},
  \citenamefont {Ritort}, \citenamefont {Jarzynski}, \citenamefont {Smith},
  \citenamefont {I.~Tinoco},\ and\ \citenamefont {Bustamante}}]{CRJe05}%
  \BibitemOpen
  \bibfield  {author} {\bibinfo {author} {\bibfnamefont {D.}~\bibnamefont
  {Collin}}, \bibinfo {author} {\bibfnamefont {F.}~\bibnamefont {Ritort}},
  \bibinfo {author} {\bibfnamefont {C.}~\bibnamefont {Jarzynski}}, \bibinfo
  {author} {\bibfnamefont {S.~B.}\ \bibnamefont {Smith}}, \bibinfo {author}
  {\bibfnamefont {J.}~\bibnamefont {I.~Tinoco}}, \ and\ \bibinfo {author}
  {\bibfnamefont {C.}~\bibnamefont {Bustamante}},\ }\href@noop {} {\bibfield
  {journal} {\bibinfo  {journal} {Nature}\ }\textbf {\bibinfo {volume} {437}},\
  \bibinfo {pages} {231} (\bibinfo {year} {2005})}\BibitemShut {NoStop}%
\bibitem [{\citenamefont {Lopez}\ \emph {et~al.}(2008)\citenamefont {Lopez},
  \citenamefont {Kuwada}, \citenamefont {Craig}, \citenamefont {Long},\ and\
  \citenamefont {Linke}}]{LBK08}%
  \BibitemOpen
  \bibfield  {author} {\bibinfo {author} {\bibfnamefont {B.~J.}\ \bibnamefont
  {Lopez}}, \bibinfo {author} {\bibfnamefont {N.~J.}\ \bibnamefont {Kuwada}},
  \bibinfo {author} {\bibfnamefont {E.~M.}\ \bibnamefont {Craig}}, \bibinfo
  {author} {\bibfnamefont {B.~R.}\ \bibnamefont {Long}}, \ and\ \bibinfo
  {author} {\bibfnamefont {H.}~\bibnamefont {Linke}},\ }\href {\doibase
  10.1103/PhysRevLett.101.220601} {\bibfield  {journal} {\bibinfo  {journal}
  {Phys. Rev. Lett.}\ }\textbf {\bibinfo {volume} {101}},\ \bibinfo {pages}
  {220601} (\bibinfo {year} {2008})}\BibitemShut {NoStop}%
\bibitem [{\citenamefont {Toyabe}\ \emph {et~al.}(2010)\citenamefont {Toyabe},
  \citenamefont {Sagawa}, \citenamefont {Ueda}, \citenamefont {Muneyuki},\ and\
  \citenamefont {Sano}}]{TSUe10}%
  \BibitemOpen
  \bibfield  {author} {\bibinfo {author} {\bibfnamefont {S.}~\bibnamefont
  {Toyabe}}, \bibinfo {author} {\bibfnamefont {T.}~\bibnamefont {Sagawa}},
  \bibinfo {author} {\bibfnamefont {M.}~\bibnamefont {Ueda}}, \bibinfo {author}
  {\bibfnamefont {E.}~\bibnamefont {Muneyuki}}, \ and\ \bibinfo {author}
  {\bibfnamefont {M.}~\bibnamefont {Sano}},\ }\href@noop {} {\bibfield
  {journal} {\bibinfo  {journal} {Nature Phys.}\ }\textbf {\bibinfo {volume}
  {6}},\ \bibinfo {pages} {988} (\bibinfo {year} {2010})}\BibitemShut {NoStop}%
\bibitem [{\citenamefont {Kim}\ and\ \citenamefont {Qian}(2007)}]{KQ07}%
  \BibitemOpen
  \bibfield  {author} {\bibinfo {author} {\bibfnamefont {K.~H.}\ \bibnamefont
  {Kim}}\ and\ \bibinfo {author} {\bibfnamefont {H.}~\bibnamefont {Qian}},\
  }\href {\doibase 10.1103/PhysRevE.75.022102} {\bibfield  {journal} {\bibinfo
  {journal} {Phys. Rev. E}\ }\textbf {\bibinfo {volume} {75}},\ \bibinfo
  {pages} {022102} (\bibinfo {year} {2007})}\BibitemShut {NoStop}%
\bibitem [{\citenamefont {Andrieux}\ and\ \citenamefont
  {Gaspard}(2008)}]{ADG08}%
  \BibitemOpen
  \bibfield  {author} {\bibinfo {author} {\bibfnamefont {D.}~\bibnamefont
  {Andrieux}}\ and\ \bibinfo {author} {\bibfnamefont {P.}~\bibnamefont
  {Gaspard}},\ }\href@noop {} {\bibfield  {journal} {\bibinfo  {journal} {Proc.
  Natl. Acad. Sci. U.S.A.}\ }\textbf {\bibinfo {volume} {105}},\ \bibinfo
  {pages} {9516} (\bibinfo {year} {2008})}\BibitemShut {NoStop}%
\bibitem [{\citenamefont {Esposito}\ and\ \citenamefont {{Van den
  Broeck}}(2011)}]{EvB11}%
  \BibitemOpen
  \bibfield  {author} {\bibinfo {author} {\bibfnamefont {M.}~\bibnamefont
  {Esposito}}\ and\ \bibinfo {author} {\bibfnamefont {C.}~\bibnamefont {{Van
  den Broeck}}},\ }\href@noop {} {\bibfield  {journal} {\bibinfo  {journal}
  {Europhys. Lett.}\ }\textbf {\bibinfo {volume} {95}},\ \bibinfo {pages}
  {40004} (\bibinfo {year} {2011})}\BibitemShut {NoStop}%
\bibitem [{\citenamefont {Sagawa}\ and\ \citenamefont {Ueda}(2008)}]{SU08}%
  \BibitemOpen
  \bibfield  {author} {\bibinfo {author} {\bibfnamefont {T.}~\bibnamefont
  {Sagawa}}\ and\ \bibinfo {author} {\bibfnamefont {M.}~\bibnamefont {Ueda}},\
  }\href@noop {} {\bibfield  {journal} {\bibinfo  {journal} {Phys. Rev. Lett.}\
  }\textbf {\bibinfo {volume} {100}},\ \bibinfo {pages} {080403} (\bibinfo
  {year} {2008})}\BibitemShut {NoStop}%
\bibitem [{\citenamefont {Sagawa}\ and\ \citenamefont {Ueda}(2009)}]{SU09}%
  \BibitemOpen
  \bibfield  {author} {\bibinfo {author} {\bibfnamefont {T.}~\bibnamefont
  {Sagawa}}\ and\ \bibinfo {author} {\bibfnamefont {M.}~\bibnamefont {Ueda}},\
  }\href@noop {} {\bibfield  {journal} {\bibinfo  {journal} {Phys. Rev. Lett.}\
  }\textbf {\bibinfo {volume} {102}},\ \bibinfo {pages} {250602} (\bibinfo
  {year} {2009})}\BibitemShut {NoStop}%
\bibitem [{\citenamefont {Sagawa}\ and\ \citenamefont {Ueda}(2010)}]{SU10}%
  \BibitemOpen
  \bibfield  {author} {\bibinfo {author} {\bibfnamefont {T.}~\bibnamefont
  {Sagawa}}\ and\ \bibinfo {author} {\bibfnamefont {M.}~\bibnamefont {Ueda}},\
  }\href@noop {} {\bibfield  {journal} {\bibinfo  {journal} {Phys. Rev. Lett.}\
  }\textbf {\bibinfo {volume} {104}},\ \bibinfo {pages} {090602} (\bibinfo
  {year} {2010})}\BibitemShut {NoStop}%
\bibitem [{\citenamefont {Horowitz}\ and\ \citenamefont
  {Vaikuntanathan}(2010)}]{HVS10}%
  \BibitemOpen
  \bibfield  {author} {\bibinfo {author} {\bibfnamefont {J.~M.}\ \bibnamefont
  {Horowitz}}\ and\ \bibinfo {author} {\bibfnamefont {S.}~\bibnamefont
  {Vaikuntanathan}},\ }\href {\doibase 10.1103/PhysRevE.82.061120} {\bibfield
  {journal} {\bibinfo  {journal} {Phys. Rev. E}\ }\textbf {\bibinfo {volume}
  {82}},\ \bibinfo {pages} {061120} (\bibinfo {year} {2010})}\BibitemShut
  {NoStop}%
\bibitem [{\citenamefont {Horowitz}\ and\ \citenamefont
  {Parrondo}(2011)}]{HP11}%
  \BibitemOpen
  \bibfield  {author} {\bibinfo {author} {\bibfnamefont {J.~M.}\ \bibnamefont
  {Horowitz}}\ and\ \bibinfo {author} {\bibfnamefont {J.~M.~R.}\ \bibnamefont
  {Parrondo}},\ }\href {http://stacks.iop.org/0295-5075/95/i=1/a=10005}
  {\bibfield  {journal} {\bibinfo  {journal} {Europhys. Lett}\ }\textbf
  {\bibinfo {volume} {95}},\ \bibinfo {pages} {10005} (\bibinfo {year}
  {2011})}\BibitemShut {NoStop}%
\bibitem [{\citenamefont {Kim}\ \emph {et~al.}(2011)\citenamefont {Kim},
  \citenamefont {Sagawa}, \citenamefont {De~Liberato},\ and\ \citenamefont
  {Ueda}}]{Kim11}%
  \BibitemOpen
  \bibfield  {author} {\bibinfo {author} {\bibfnamefont {S.~W.}\ \bibnamefont
  {Kim}}, \bibinfo {author} {\bibfnamefont {T.}~\bibnamefont {Sagawa}},
  \bibinfo {author} {\bibfnamefont {S.}~\bibnamefont {De~Liberato}}, \ and\
  \bibinfo {author} {\bibfnamefont {M.}~\bibnamefont {Ueda}},\ }\href {\doibase
  10.1103/PhysRevLett.106.070401} {\bibfield  {journal} {\bibinfo  {journal}
  {Phys. Rev. Lett.}\ }\textbf {\bibinfo {volume} {106}},\ \bibinfo {pages}
  {070401} (\bibinfo {year} {2011})}\BibitemShut {NoStop}%
\bibitem [{\citenamefont {Abreu}\ and\ \citenamefont {Seifert}(2011)}]{AS11}%
  \BibitemOpen
  \bibfield  {author} {\bibinfo {author} {\bibfnamefont {D.}~\bibnamefont
  {Abreu}}\ and\ \bibinfo {author} {\bibfnamefont {U.}~\bibnamefont
  {Seifert}},\ }\href {http://stacks.iop.org/0295-5075/94/i=1/a=10001}
  {\bibfield  {journal} {\bibinfo  {journal} {Europhys. Lett}\ }\textbf
  {\bibinfo {volume} {94}},\ \bibinfo {pages} {10001} (\bibinfo {year}
  {2011})}\BibitemShut {NoStop}%
\bibitem [{\citenamefont {Abreu}\ and\ \citenamefont {Seifert}(2012)}]{AS12}%
  \BibitemOpen
  \bibfield  {author} {\bibinfo {author} {\bibfnamefont {D.}~\bibnamefont
  {Abreu}}\ and\ \bibinfo {author} {\bibfnamefont {U.}~\bibnamefont
  {Seifert}},\ }\href {\doibase 10.1103/PhysRevLett.108.030601} {\bibfield
  {journal} {\bibinfo  {journal} {Phys. Rev. Lett.}\ }\textbf {\bibinfo
  {volume} {108}},\ \bibinfo {pages} {030601} (\bibinfo {year}
  {2012})}\BibitemShut {NoStop}%
\bibitem [{\citenamefont {Esposito}\ and\ \citenamefont
  {Schaller}(2012)}]{ES12}%
  \BibitemOpen
  \bibfield  {author} {\bibinfo {author} {\bibfnamefont {M.}~\bibnamefont
  {Esposito}}\ and\ \bibinfo {author} {\bibfnamefont {G.}~\bibnamefont
  {Schaller}},\ }\href {http://stacks.iop.org/0295-5075/99/i=3/a=30003}
  {\bibfield  {journal} {\bibinfo  {journal} {Europhys. Lett}\ }\textbf
  {\bibinfo {volume} {99}},\ \bibinfo {pages} {30003} (\bibinfo {year}
  {2012})}\BibitemShut {NoStop}%
\bibitem [{\citenamefont {Sagawa}\ and\ \citenamefont {Ueda}(2012)}]{SU12}%
  \BibitemOpen
  \bibfield  {author} {\bibinfo {author} {\bibfnamefont {T.}~\bibnamefont
  {Sagawa}}\ and\ \bibinfo {author} {\bibfnamefont {M.}~\bibnamefont {Ueda}},\
  }\href {\doibase 10.1103/PhysRevE.85.021104} {\bibfield  {journal} {\bibinfo
  {journal} {Phys. Rev. E}\ }\textbf {\bibinfo {volume} {85}},\ \bibinfo
  {pages} {021104} (\bibinfo {year} {2012})}\BibitemShut {NoStop}%
\bibitem [{\citenamefont {Maxwell}(1871)}]{Max1871}%
  \BibitemOpen
  \bibfield  {author} {\bibinfo {author} {\bibfnamefont {J.~C.}\ \bibnamefont
  {Maxwell}},\ }\href@noop {} {\emph {\bibinfo {title} {Theory of Heat}}}\
  (\bibinfo  {publisher} {Appleton},\ \bibinfo {year} {1871})\BibitemShut
  {NoStop}%
\bibitem [{\citenamefont {Leff}\ and\ \citenamefont {Rex}(2010)}]{LM10}%
  \BibitemOpen
  \bibfield  {author} {\bibinfo {author} {\bibfnamefont {H.}~\bibnamefont
  {Leff}}\ and\ \bibinfo {author} {\bibfnamefont {A.}~\bibnamefont {Rex}},\
  }\href {http://books.google.co.jp/books?id=VNKCsQt75\_UC} {\emph {\bibinfo
  {title} {Maxwell's Demon 2 Entropy, Classical and Quantum Information,
  Computing}}},\ Maxwell's Demon\ (\bibinfo  {publisher} {Taylor \& Francis},\
  \bibinfo {year} {2010})\BibitemShut {NoStop}%
\bibitem [{\citenamefont {Maruyama}\ \emph {et~al.}(2009)\citenamefont
  {Maruyama}, \citenamefont {Nori},\ and\ \citenamefont {Vedral}}]{MNV09}%
  \BibitemOpen
  \bibfield  {author} {\bibinfo {author} {\bibfnamefont {K.}~\bibnamefont
  {Maruyama}}, \bibinfo {author} {\bibfnamefont {F.}~\bibnamefont {Nori}}, \
  and\ \bibinfo {author} {\bibfnamefont {V.}~\bibnamefont {Vedral}},\ }\href
  {\doibase 10.1103/RevModPhys.81.1} {\bibfield  {journal} {\bibinfo  {journal}
  {Rev. Mod. Phys.}\ }\textbf {\bibinfo {volume} {81}},\ \bibinfo {pages} {1}
  (\bibinfo {year} {2009})}\BibitemShut {NoStop}%
\bibitem [{\citenamefont {Jacobs}(2009)}]{Ja09}%
  \BibitemOpen
  \bibfield  {author} {\bibinfo {author} {\bibfnamefont {K.}~\bibnamefont
  {Jacobs}},\ }\href {\doibase 10.1103/PhysRevA.80.012322} {\bibfield
  {journal} {\bibinfo  {journal} {Phys. Rev. A}\ }\textbf {\bibinfo {volume}
  {80}},\ \bibinfo {pages} {012322} (\bibinfo {year} {2009})}\BibitemShut
  {NoStop}%
\bibitem [{\citenamefont {Hasegawa}\ \emph {et~al.}(2010)\citenamefont
  {Hasegawa}, \citenamefont {Ishikawa}, \citenamefont {Takara},\ and\
  \citenamefont {Driebe}}]{HITD10}%
  \BibitemOpen
  \bibfield  {author} {\bibinfo {author} {\bibfnamefont {H.-H.}\ \bibnamefont
  {Hasegawa}}, \bibinfo {author} {\bibfnamefont {J.}~\bibnamefont {Ishikawa}},
  \bibinfo {author} {\bibfnamefont {K.}~\bibnamefont {Takara}}, \ and\ \bibinfo
  {author} {\bibfnamefont {D.}~\bibnamefont {Driebe}},\ }\href {\doibase
  http://dx.doi.org/10.1016/j.physleta.2009.12.042} {\bibfield  {journal}
  {\bibinfo  {journal} {Phys. Lett. A}\ }\textbf {\bibinfo {volume} {374}},\
  \bibinfo {pages} {1001 } (\bibinfo {year} {2010})}\BibitemShut {NoStop}%
\bibitem [{\citenamefont {Takara}\ \emph {et~al.}(2010)\citenamefont {Takara},
  \citenamefont {Hasegawa},\ and\ \citenamefont {Driebe}}]{THD10}%
  \BibitemOpen
  \bibfield  {author} {\bibinfo {author} {\bibfnamefont {K.}~\bibnamefont
  {Takara}}, \bibinfo {author} {\bibfnamefont {H.-H.}\ \bibnamefont
  {Hasegawa}}, \ and\ \bibinfo {author} {\bibfnamefont {D.}~\bibnamefont
  {Driebe}},\ }\href {\doibase
  http://dx.doi.org/10.1016/j.physleta.2010.11.002} {\bibfield  {journal}
  {\bibinfo  {journal} {Phys. Lett. A}\ }\textbf {\bibinfo {volume} {375}},\
  \bibinfo {pages} {88 } (\bibinfo {year} {2010})}\BibitemShut {NoStop}%
\bibitem [{\citenamefont {{Murashita}}\ \emph {et~al.}()\citenamefont
  {{Murashita}}, \citenamefont {{Funo}},\ and\ \citenamefont {{Ueda}}}]{MFU14}%
  \BibitemOpen
  \bibfield  {author} {\bibinfo {author} {\bibfnamefont {Y.}~\bibnamefont
  {{Murashita}}}, \bibinfo {author} {\bibfnamefont {K.}~\bibnamefont {{Funo}}},
  \ and\ \bibinfo {author} {\bibfnamefont {M.}~\bibnamefont {{Ueda}}},\
  }\href@noop {} {\ }\Eprint {http://arxiv.org/abs/1401.4494} {arXiv:1401.4494
  [cond-mat.stat-mech]} \BibitemShut {NoStop}%
\bibitem [{\citenamefont {Lahiri}\ \emph {et~al.}(2012)\citenamefont {Lahiri},
  \citenamefont {Rana},\ and\ \citenamefont {Jayannavar}}]{SJ12}%
  \BibitemOpen
  \bibfield  {author} {\bibinfo {author} {\bibfnamefont {S.}~\bibnamefont
  {Lahiri}}, \bibinfo {author} {\bibfnamefont {S.}~\bibnamefont {Rana}}, \ and\
  \bibinfo {author} {\bibfnamefont {A.~M.}\ \bibnamefont {Jayannavar}},\ }\href
  {http://stacks.iop.org/1751-8121/45/i=6/a=065002} {\bibfield  {journal}
  {\bibinfo  {journal} {J. Phys. A: Math. Theor.}\ }\textbf {\bibinfo {volume}
  {45}},\ \bibinfo {pages} {065002} (\bibinfo {year} {2012})}\BibitemShut
  {NoStop}%
\bibitem [{\citenamefont {Harris}\ and\ \citenamefont
  {Sch{\"u}tz}(2007)}]{HS07}%
  \BibitemOpen
  \bibfield  {author} {\bibinfo {author} {\bibfnamefont {R.~J.}\ \bibnamefont
  {Harris}}\ and\ \bibinfo {author} {\bibfnamefont {G.~M.}\ \bibnamefont
  {Sch{\"u}tz}},\ }\href {http://stacks.iop.org/1742-5468/2007/i=07/a=P07020}
  {\bibfield  {journal} {\bibinfo  {journal} {J. Stat. Mech.}\ }\textbf
  {\bibinfo {volume} {2007}},\ \bibinfo {pages} {P07020} (\bibinfo {year}
  {2007})}\BibitemShut {NoStop}%
\bibitem [{\citenamefont {Seifert}(2012)}]{Sei12}%
  \BibitemOpen
  \bibfield  {author} {\bibinfo {author} {\bibfnamefont {U.}~\bibnamefont
  {Seifert}},\ }\href@noop {} {\bibfield  {journal} {\bibinfo  {journal} {Rep.
  Prog. Phys.}\ }\textbf {\bibinfo {volume} {75}},\ \bibinfo {pages} {126001}
  (\bibinfo {year} {2012})}\BibitemShut {NoStop}%
\bibitem [{\citenamefont {Esposito}\ and\ \citenamefont {Van~den
  Broeck}(2010)}]{EB10}%
  \BibitemOpen
  \bibfield  {author} {\bibinfo {author} {\bibfnamefont {M.}~\bibnamefont
  {Esposito}}\ and\ \bibinfo {author} {\bibfnamefont {C.}~\bibnamefont {Van~den
  Broeck}},\ }\href {\doibase 10.1103/PhysRevLett.104.090601} {\bibfield
  {journal} {\bibinfo  {journal} {Phys. Rev. Lett.}\ }\textbf {\bibinfo
  {volume} {104}},\ \bibinfo {pages} {090601} (\bibinfo {year}
  {2010})}\BibitemShut {NoStop}%
\bibitem [{\citenamefont {Seifert}(2005)}]{Sei05}%
  \BibitemOpen
  \bibfield  {author} {\bibinfo {author} {\bibfnamefont {U.}~\bibnamefont
  {Seifert}},\ }\href {\doibase 10.1103/PhysRevLett.95.040602} {\bibfield
  {journal} {\bibinfo  {journal} {Phys. Rev. Lett.}\ }\textbf {\bibinfo
  {volume} {95}},\ \bibinfo {pages} {040602} (\bibinfo {year}
  {2005})}\BibitemShut {NoStop}%
\bibitem [{\citenamefont {Szilard}(1929)}]{Szi29}%
  \BibitemOpen
  \bibfield  {author} {\bibinfo {author} {\bibfnamefont {L.}~\bibnamefont
  {Szilard}},\ }\href@noop {} {\bibfield  {journal} {\bibinfo  {journal} {Z.
  Phys.}\ }\textbf {\bibinfo {volume} {53}},\ \bibinfo {pages} {840} (\bibinfo
  {year} {1929})}\BibitemShut {NoStop}%
\bibitem [{\citenamefont {Plesch}\ \emph {et~al.}(2013)\citenamefont {Plesch},
  \citenamefont {Dahlsten}, \citenamefont {Goold},\ and\ \citenamefont
  {Vedral}}]{POV13}%
  \BibitemOpen
  \bibfield  {author} {\bibinfo {author} {\bibfnamefont {M.}~\bibnamefont
  {Plesch}}, \bibinfo {author} {\bibfnamefont {O.}~\bibnamefont {Dahlsten}},
  \bibinfo {author} {\bibfnamefont {J.}~\bibnamefont {Goold}}, \ and\ \bibinfo
  {author} {\bibfnamefont {V.}~\bibnamefont {Vedral}},\ }\href {\doibase
  10.1103/PhysRevLett.111.188901} {\bibfield  {journal} {\bibinfo  {journal}
  {Phys. Rev. Lett.}\ }\textbf {\bibinfo {volume} {111}},\ \bibinfo {pages}
  {188901} (\bibinfo {year} {2013})}\BibitemShut {NoStop}%
\bibitem [{\citenamefont {Kim}\ \emph {et~al.}(2013)\citenamefont {Kim},
  \citenamefont {Kim}, \citenamefont {Sagawa}, \citenamefont {De~Liberato},\
  and\ \citenamefont {Ueda}}]{Kim13}%
  \BibitemOpen
  \bibfield  {author} {\bibinfo {author} {\bibfnamefont {S.~W.}\ \bibnamefont
  {Kim}}, \bibinfo {author} {\bibfnamefont {K.-H.}\ \bibnamefont {Kim}},
  \bibinfo {author} {\bibfnamefont {T.}~\bibnamefont {Sagawa}}, \bibinfo
  {author} {\bibfnamefont {S.}~\bibnamefont {De~Liberato}}, \ and\ \bibinfo
  {author} {\bibfnamefont {M.}~\bibnamefont {Ueda}},\ }\href {\doibase
  10.1103/PhysRevLett.111.188902} {\bibfield  {journal} {\bibinfo  {journal}
  {Phys. Rev. Lett.}\ }\textbf {\bibinfo {volume} {111}},\ \bibinfo {pages}
  {188902} (\bibinfo {year} {2013})}\BibitemShut {NoStop}%
\bibitem [{\citenamefont {Cover}\ and\ \citenamefont {Thomas}(2006)}]{Cov06}%
  \BibitemOpen
  \bibfield  {author} {\bibinfo {author} {\bibfnamefont {T.~M.}\ \bibnamefont
  {Cover}}\ and\ \bibinfo {author} {\bibfnamefont {J.~A.}\ \bibnamefont
  {Thomas}},\ }\href@noop {} {\emph {\bibinfo {title} {Elements of Information
  Theory (Wiley Series in Telecommunications and Signal Processing)}}}\
  (\bibinfo  {publisher} {Wiley-Interscience},\ \bibinfo {year}
  {2006})\BibitemShut {NoStop}%
\bibitem [{TY()}]{TY}%
  \BibitemOpen
  \href@noop {} {}\bibinfo {note} {S. Toyabe (private
  communication).}\BibitemShut {Stop}%
\end{thebibliography}%

\end{document}